\documentclass[prb,twocolumn,amsmath,amssymb,showpacs]{revtex4-1}
\usepackage{graphicx}

\def\beq{\begin{eqnarray}}
\def\eeq{\end{eqnarray}}
\def\beqa{\begin{eqnarray}}
\def\eeqa{\end{eqnarray}}

\begin{document}

\title{Doping and temperature dependence of the pseudogap and Fermi arcs in 
cuprates from $d$-CDW short-range fluctuations in the context of the $t$-$J$ model
}

\author{Mat\'{\i}as Bejas, Guillermo Buzon, Andr\'es Greco, and Adriana Foussats}

\affiliation{
Facultad de Ciencias Exactas, Ingenier\'{\i}a y Agrimensura and
Instituto de F\'{\i}sica Rosario, Universidad Nacional de Rosario and Consejo Nacional de Investigaciones Cient\'{\i}ficas y T\'ecnicas,
Avenida Pellegrini 250-2000 Rosario-Argentina.
}

\date{\today}

\begin{abstract}
At mean-field level the $t$-$J$  model shows a phase diagram with close 
analogies to the phase diagram of hole doped cuprates.
An order parameter associated with the flux
or $d$ charge-density wave ($d$-CDW) phase competes and coexists with superconductivity at low 
doping showing characteristics identified with the observed pseudogap
in underdoped cuprates.
In addition, in the $d$-CDW state the Fermi surface is 
reconstructed toward pockets with low spectral weight in the outer part, 
resembling the arcs observed in angle-resolved  photoemission spectroscopy
experiments.
However, the $d$-CDW requires broken translational symmetry,
a fact that is not completely accepted. 
Including self-energy corrections beyond the mean, field we found 
that the self-energy can be written as two distinct contributions.
One of these (called $\Sigma _{flux}$) dominates at low 
energy and originates from the scattering between carriers and $d$-CDW 
fluctuations in proximity to the $d$-CDW instability.
The second contribution (called $\Sigma_{R\lambda}$) dominates at large energy and
originates from the scattering between charge fluctuations under the 
constraint of non double occupancy.   
In this paper it is shown that $\Sigma_{flux}$ is responsible for 
the origin of low-energy features  in the spectral function as 
a pseudogap and Fermi arcs.
The obtained doping and temperature dependence of the 
pseudogap and Fermi arcs is similar to that observed in experiments.
At low energy, $\Sigma_{R \lambda}$ gives an additional contribution to the 
closure of the pseudogap. 
\end{abstract}

\maketitle

\section{Introduction}

The origin of the pseudogap (PG) phase in cuprates is one of the most important and unresolved issues in solid-state physics.\cite{timusk99} 
Several experimental techniques are used for studying this phase,
and its main characteristics remain unclear.
For instance, in the  superconducting state, some experiments are consistent with 
the existence of only one gap while others are in 
agreement with two order (competing) parameters.
Two main scenarios, preformed pairs above $T_c$ and two competing 
gaps, dispute the explanation of the PG (see Refs. \onlinecite{hufner08} and \onlinecite{norman05}).

Angle-resolved photoemission spectroscopy (ARPES) is presently a valuable tool for such research.\cite{damascelli03}
Surprisingly, in underdoped cuprates, ARPES shows, in the normal state below a characteristic temperature $T^*$,   
Fermi arcs\cite{norman98,kanigel06,kanigel07,shi,terashima07,hashimoto10,yoshida09,kondo07,kondo09,lee07,tanaka06,ma08} (FAs) 
(centered along the zone diagonal) instead of the full Fermi surface (FS) 
predicted by standard solid-state physics.
Despite the general consensus about the existence of FAs, their main characteristics are controversial, even 
at an experimental level. For instance, while some experiments suggest that the arcs are disconnected,\cite{norman98,kanigel06,kanigel07}
others claim that the arcs are associated with pockets.\cite{Nd,meng,jonson1,jonson2,jonson3}

The number of theoretical studies about the PG and FAs is too large for a complete listing. 
Among them, early results where the PG was 
discussed in the framework of the Born approximation and the spin-polaron description for  the $t$-$J$ model should be mentioned.\cite{sherman97}
In addition, recent progress on dynamical cluster approximation (DCA) show the presence of a pseudogap \cite{huscroft01,macridin06} and 
Fermi arcs\cite{parcollet04,werner09,lin10} in the two-dimensional Hubbard model.

Recently, Norman {\it et al.} (Ref. \onlinecite{norman07}) have summarized some  
of the relevant models proposed for discussing ARPES experiments. 
These models are semiphenomenological or phenomenological, and the proposed Green function $G(k,\omega)$ 
has the following form:
\begin{eqnarray}
G^{-1}(k,\omega)=\omega-\epsilon_k + i \Gamma - \Sigma(k,\omega)
\end{eqnarray}
\noindent where $\Gamma$ is a lifetime broadening, $\epsilon_k$ is the bare electronic dispersion, and $\Sigma(k,\omega)$ is the self-energy, which 
can be written as
\begin{eqnarray}
\Sigma(k,\omega) = \frac{\Delta^2_k}{\omega + \xi_k + i\Gamma}
\end{eqnarray}
\noindent In Eq. (2) the phenomenological pseudogap $\Delta_k$ is assumed to be $d$ wave;
$\Delta_k = \frac{\Delta}{2} (\cos k_x - \cos k_y)$.
$\xi_k$ is model dependent; for instance:
(a) $\xi_k = -\epsilon_{k+Q}$, where $Q=(\pi,\pi)$, in the $d$ charge-density wave ($d$-CDW) model,\cite{chakravarty01}
(b) $\xi_k$ is the nearest-neighbor term of the tight binding dispersion in the model proposed by Yang, Rice, 
and Zhang (YRZ),\cite{yang06}
and (c) $\xi_k = \epsilon_k$ in the $d$-wave preformed pairs model.\cite{norman07,norman98p}
Although these models represent different physical situations, the experimental distinction
between them is a big challenge.

In Ref. \onlinecite{norman07} it was concluded that $d$-CDW and  YRZ models lead to predictions that are not compatible with experiments.
For instance, these models lead to FAs whose length is temperature independent and deviates  from the underlying FS in contrast to the 
experiments.
Finally, it was also concluded that experiments are better described in the framework of the $d$-wave pairs model.

Since the early studies on high-Tc superconductivity, the $t$-$J$ model has been shown to be a basic model for describing the physics 
of cuprates.\cite{anderson87}
This model, which can be considered a strong-coupling version of the Hubbard model,\cite{ZR,Oles} contains (potentially) 
the main ingredients for describing cuprates, i.e., antiferromagnetism at zero doping, a metallic phase at 
finite doping, strong tendency to $d$-wave superconductivity
and several candidates for the PG phase at low doping.
Whether all these phases may be unambiguously associated to those known in cuprates is the big challenge
for the $t$-$J$ model.
The number of analytical and numerical techniques introduced for studying this model is too large
to discuss here.

One analytical approach for treating the model is the large-$N$ expansion where the two spin components are 
extended to $N$ and an expansion in powers of the small parameter $1/N$ is performed.
The advantage of this approach is that the results are not perturbative in any model parameter and occur in strong coupling.
For performing the large-$N$ expansion, treatments based on the slave boson \cite{wang92} and Hubbard operators\cite{zeyher2} were developed.
However, evaluating fluctuations above mean field, as required for calculating dynamical self-energy effects, 
is not straightforward.\cite{wang92}
On the basis of the path-integral representation for Hubbard operators\cite{foussats04} the large-$N$ approach to the $t$-$J$ model 
was implemented, yielding previous results\cite{wang92,zeyher2} in leading order. 
At mean-field level ($N=\infty$) the well-known flux phase\cite{affleck88,kotliar88,morse91,cappelluti99} (FP) 
instability at low doping was also reobtained.\cite{foussats04}
In the FP a charge-density wave coexists with orbital 
currents in a staggered pattern and has the same momentum dependence of the
superconducting state ($d$ wave), allowing the identification
of the FP with the pseudogap. In addition, the FP scenario possesses the main properties to be identified with the phenomenological 
$d$-CDW  proposal.\cite{chakravarty01} 
It is important to mention that the relevance of the FP for the physical case $N=2$, for instance, in the form of a phase with strong $d$-wave 
short-range order, is under dispute. While some exact diagonalization results\cite{leung00} show the presence of 
the $d$-CDW phase, DCA (Ref. \onlinecite{macridin04}) and strong-coupling diagram technique\cite{sherman08} do not show the static
long-range formation of the $d$-CDW state. In spite of this discussion it is important to note that  the 
predicted mean-field phase diagram\cite{cappelluti99} 
has close similarities to the phase diagram of hole-doped cuprates
where the FP competes and coexists with superconductivity.

Importantly, the method developed in Ref. \onlinecite{foussats04} allows us to go beyond the mean field and to compute self-energy 
renormalizations.
Here, following Refs. \onlinecite{greco08} and \onlinecite{greco09}, it will be shown that the 
doping and temperature dependence of the PG and FAs can be discussed after
including self-energy effects in proximity to the FP instability, showing that $d$-CDW model is not 
inconsistent with the notion of arcs.

This paper is organized as follows.
In Sec. II we summarize the basic formalism.
We show that the self-energy can be written in terms of two distinct contributions, $\Sigma_{flux}$ and $\Sigma_{R\lambda}$. 
The mean-field phase diagram is discussed together 
with the main characteristics of the self-energy.
In Sec. III we describe the origin of the FAs and show that they are triggered by $\Sigma_{flux}$. 
Section III A discusses the topology of the FAs.
Sections III B and III C discuss the temperature 
and doping dependence of the FAs, respectively. In Sec. III D we present the 
main characteristics of $\Sigma_{flux}$ at finite temperature. 
In Sec. IV we discuss the inclussion of $\Sigma_{R\lambda}$. 
Section V presents discussion and conclusions.

\section{Basic framework}

The large-$N$ mean-field solution of the $t$-$J$ model yields a quasiparticle (QP) dispersion:\cite{foussats04}
\begin{eqnarray}
\epsilon_{k} &=& -2\left( t \frac{\delta}{2} + rJ \right) [cos(k_x)+cos(k_y)] \nonumber \\
             & & -4 \: t'\: \frac{\delta}{2} \; cos(k_x) \: cos(k_y) - \mu,
\end{eqnarray}
\noindent where  $\delta$ is the doping away from half-filling.
$t$, $t'$, and $J$ are hopping between nearest-neighbor, next-nearest-neighbor, and the nearest-neighbor Heisenberg coupling,
respectively.
The contribution  
$r$ to the mean-field band and the chemical potential $\mu$ must be obtained self-consistently from  
\begin{eqnarray}
r=\frac{1}{N_s} \sum_{k} cos(k_x) n_F(\epsilon_{k})
\end{eqnarray}
and
\begin{eqnarray}
(1-\delta)=\frac{1}{N_s} \sum_{k} n_F(\epsilon_{k}),
\end{eqnarray}
where $n_F$ is the Fermi factor and $N_s$ the number of sites.

Equations (3)–(5) define a homogeneous Fermi liquid (HFL) phase that, as discussed in Sec. I, is unstable against a flux phase or $d$-CDW 
state at low doping.

Beyond the mean field the computation of fluctuations in $O(1/N)$ leads to the following expression for the self-energy:\cite{bejas06}
\begin{eqnarray}\label{SigmaIm}
 {\mathrm{Im}}\Sigma(k,\omega) = -\frac{1}{N_{s}}
\sum_{q,a,b} h_{a}(k,q,\omega-\epsilon_{{k-q}}) h_{b}(k,q,\omega-\epsilon_{{k-q}}) \nonumber \\ 
\times {\mathrm{Im}}[D_{ab}(q,\omega-\epsilon_{{k-q}})]
 [n_{F}( -\epsilon_{k-q}) +n_{B}(\omega-\epsilon_{{k-q}})] \nonumber \\
\end{eqnarray}
\noindent where $n_B$ is the Bose factor and the six-component vector $h_{a} (k,q,\nu)$ is
\begin{widetext}
\begin{eqnarray}
h_{a} (k,q,\nu)
&=&\left\{ \frac{}{} \right. \frac{2\epsilon_{k-q}+\nu+2\mu}{2}
  + J r \left[ \cos\left(k_x-\frac{q_x}{2} \right) \cos \left( \frac{q_x}{2} \right) +
\cos\left(k_y-\frac{q_y}{2} \right) \cos \left( \frac{q_y}{2} \right) \right]
 \; ; 1 \; ;\nonumber \\
&& -J r \; \cos \left( k_{x}-\frac{q_{x}}{2} \right) \; ;
-J r \; \cos \left( k_{y}-\frac{q_{y}}{2} \right) \; ; \; 
 J r \; \sin \left( k_{x}-\frac{q_{x}}{2} \right) \; ; \;
J r \; \sin \left( k_{y}-\frac{q_{y}}{2} \right) \left. \frac{}{} \right\}.
\end{eqnarray}
\end{widetext}

The physical information contained in the vector $h_{a} (k,q,\nu)$ is as follows. 
The first component (called $\delta R$) is mainly dominated by the usual charge channel, the second component 
(called $\delta \lambda$) corresponds to 
the nondouble-occupancy constraint, and the last four components are driven by $J$.
For the case $J=0$ the vector $h_a$ reduces to a two-component vector.

In Eq. (6) $D_{ab}$ is a $6\times6$ matrix that contains contributions from 
the six different channels and their mixing.

\begin{equation}
D^{-1}_{ab}(q,i\omega_{n})=[D^{(0)}_{ab}(q,i\omega_{n})]^{-1}- \Pi_{ab}(q,i\omega_{n})
\end{equation}
\noindent where

\begin{widetext}
\begin{equation} \label{eq:D0-1}
D^{(0)}_{ab}(q,i\omega_{n}) = 
\left(
  \begin{array}{cccccc}
 \delta^2/2 (V-J/2) [\cos(q_x)+\cos(q_y)]      & \delta/2   & 0         & 0       & 0     & 0 \\
   \delta/2     & 0            & 0                         & 0       & 0     & 0 \\
  0            & 0            & J\;r^2                   & 0       & 0     & 0 \\
  0            & 0            & 0                         & J\;r^2   & 0     & 0 \\
  0            & 0            & 0                         & 0         & J\;r^2   & 0   \\
  0            & 0            & 0                         & 0          & 0     & J\;r^2  \\
 \end{array}
\right)^{-1}
\end{equation}
\noindent and
\begin{equation}
\Pi_{ab}(q, i\omega_{n})
=- \frac{1}{N_s} \; \sum_{k}\;h_{a}(k,q,\epsilon_k-\epsilon_{k-q}) \;h_{b}(k,q,\epsilon_k-\epsilon_{k-q})
\; g(k,q,i\omega_{n}) 
- \delta_{a}^R\delta_{b}^R\;\frac{1}{N_s} \sum_{k} \frac{\epsilon_{k-q} - \epsilon_{k}}{2} \;
n_{F}(\epsilon_{k}) \; , 
\end{equation}
\end{widetext}

\noindent with 
\begin{eqnarray} \label{eq:gPi}
g(k,q,i\omega_{n}) = \frac{[n_{F}(\epsilon_{k - q}) - n_{F}(\epsilon_{k})]} 
{i\omega_{n} + \epsilon_{k - q} - \epsilon_{k} } \; ,
\end{eqnarray}
\noindent where $i\omega_n$ is the bosonic Matsubara frequency.
Hereafter, $t'/t=-0.35$ and $J/t=0.3$, which are suitable parameters for cuprates, are used.
The lattice constant $a$ of the square lattice and $t$ are considered to be a length unit and energy unit, respectively.
In Eq.(\ref{eq:D0-1}), $V$ is the nearest-neighbor Coulomb repulsion.
The role of $V$ is to exclude phase separation. We choose $V=2J$.

The instability of the mean-field solution occurs when $det[ D_{ab}(q,i\omega_n=0) ]=0$ (Ref. \onlinecite{foussats04}). 
For the present parameters, at $T=0$, the instability takes place at 
$\delta=\delta_c\sim 0.23$ for $q\sim (\pi,\pi)$. 
It is important to note that $D_{ab}$ enters explicitly in the self-energy expression beyond the mean field [Eq.(6)]; thus, 
$\Sigma$ probes the proximity to the instability at low $\omega$ and for momenta $k-q$ near the FS. 

Since $D_{ab}$ contains contributions from six different channels and their mixing, it is important  to isolate the 
most relevant channel dominating $\Sigma$ near the instability. The eigenvector with  zero eigenvalue of $D_{ab}$ takes the 
form $\sim (0,0,0,0,-1,1)$ which is the eigenvector associated to the FP 
instability.\cite{foussats04}
Projecting $\Sigma(k,\omega)$ on the FP eigenvector the following self-energy contribution is obtained\cite{greco08,greco09}
\begin{eqnarray}
\label{eq:SigmaIm0flux}
{\mathrm{Im}} \, \Sigma_{flux}(k,\omega) &=& 
-\frac{1}{N_{s}} \sum_{q} \gamma^2(q,k)
{\mathrm{Im}} \chi_{flux}(q,\omega-\epsilon_{{k-q}}) \nonumber \\
&\times& \left[n_{F}(-\epsilon_{{k-q}}) + n_{B}(\omega-\epsilon_{{k-q}})\right]
\end{eqnarray}
which shows the explicit contribution of the flux susceptibility
\begin{eqnarray}
\chi_{flux}(q,\omega)= [2J\;r^2-\Pi(q,\omega)]^{-1}
\end{eqnarray}
where $\Pi(q,\omega)$ is an electronic polarizability  
\begin{eqnarray}
\Pi(q, i\omega_n) = - \frac{1}{N_{s}}\;
\sum_{k}\; \gamma^2(q,k) \frac{[n_{F}(\epsilon_{k + q}) 
- n_{F}(\epsilon_{k})]} 
{\epsilon_{k + q} - \epsilon_{k}-i \omega_n}\; \nonumber \\ 
\end{eqnarray}
calculated with a form factor 
$\gamma(q,k)=2 r [\sin(k_x-q_x/2)-\sin(k_y-q_y/2)]$.
Since the instability takes place at  
$(\pi,\pi)$ the form factor $\gamma(q,k)$ transforms 
into $\sim [\cos(k_x)-\cos(k_y)]$, which indicates the $d$-wave character of the FP. Thus, the mode associated with the FP instability plays an 
important role in $\Sigma(k,\omega)$ at low doping near 
$\delta_c$. 
\begin{figure}
\includegraphics[bb=24 50 706 560,angle=0,width=8cm]{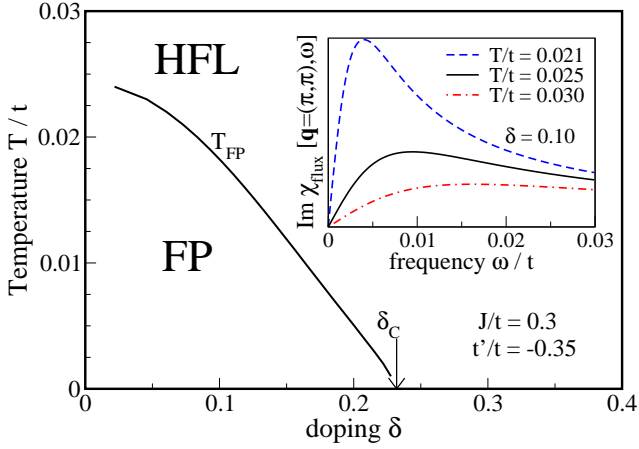}
\label{fig:intkTdp}
\caption{(Color online) Phase diagram of the $t$-$J$ model in the leading order of $1/N$ expansion where 
superconductivity was discarded.  
The instability (solid line), marked by $T_{FP}$, separates the homogeneous Fermi liquid state from the flux
or $d$-CDW state and terminates at
the quantum critical point at $\delta_{c} \sim 0.23$ at $T=0$. 
The inset shows the imaginary part of the flux susceptibility  
vs. $\omega$  for $\delta=0.10$
and for different temperatures [$q=(\pi,\pi)$ is the momentum where the instability
occurs].  Approaching $T_{FP}$ from above the flux mode becomes better 
defined, accumulates weight, and softens. At $T=T_{FP}$ the flux mode reaches 
$\omega=0$, freezing the $d$-CDW phase. This flux mode contributes to the self-energy leading to a pseudogap and Fermi 
arcs features as discussed in text.}
\end{figure}

In Fig. 1, disregarding superconductivity, the solid line shows the temperature 
$T_{FP}$, which indicates the onset of FP 
instability, i.e., when the static ($i\omega_n=0$) flux susceptibility [Eq.(13)] diverges.
At $T=0$ a phase transition occurs at the quantum critical point (QCP) placed at the critical doping $\delta_c$. 
At $T_{FP}$ a flux-mode [${\rm Im} \chi_{flux}(q=(\pi,\pi),\omega)$] reaches $\omega=0$, freezing the FP.
In the inset in Fig. 1, we have plotted ${\rm Im} \chi_{flux}(q=(\pi,\pi),\omega)$ for $\delta=0.10$ for several temperatures approaching 
$T_{FP}/t \sim 0.018$, 
showing that when $T \rightarrow T_{FP}$, a low energy $d$-wave flux mode becomes soft and accumulates large
spectral weight.
We have used the small broadening $\eta/t=0.01$ in the analytic
continuation ($i\omega_n \rightarrow \omega + i\eta$).

Figure 2(a) shows  $\rm-Im \, \Sigma_{flux}(k,\omega)$ at $T=0$ for several dopings at the antinodal Fermi wave vector $k_F^{AN}$. 
At large doping $\delta=0.40$, $\rm-Im \Sigma_{flux}$ is weak and behaves as $\sim \omega^2$ at low energy. Approaching $\delta_c$ 
($\delta=0.26$ and $\delta=0.24$), $\rm-Im \Sigma_{flux}$ 
increases, and the behavior at low energy is nearly linear in $\omega$ and 
develops structures at low energy $\omega/t \sim 0.1-0.2$.
Results for the nodal Fermi vector $k_F^N$ are not shown because they 
are nearly negligible due to the $d$-wave character of the flux instability.
Inset (i) in Fig. 2(b) shows the QP weight $Z$ at $k_F^{AN}$ ($Z_{flux}^{AN}$) and at $k_F^N$  ($Z_{flux}^{N}$). 
While $Z_{flux}^{AN}$ is strongly doping dependent and tends to zero approaching 
$\delta_c$, $Z_{flux}^N \sim 1$ shows that $\Sigma_{flux}$ is also strongly anisotropic on the FS.

$\Sigma_{flux}$ is written in terms of the flux or $d$-CDW susceptibility $\chi_{flux}(q,\omega)$,
which shows explicitly the role played by the soft flux mode with momentum $(\pi,\pi)$ (see inset in Fig. 1).
Therefore, near the antinode the QP on the FS is strongly distorted, leading to FA effects, as shown in Sec. III
In addition, it is easy to check that the most important 
$J$ contribution to $\Sigma(k,\omega)$ enters only via $\Sigma_{flux}$.

\begin{figure}
\centering
\includegraphics[width=8cm]{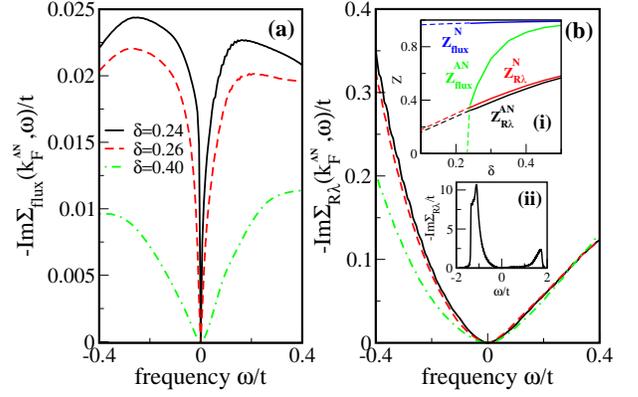}
\label{fig:imag}
\caption{(Color online)
The imaginary part of the two self-energy contributions, $\Sigma_{flux}$ and $\Sigma_{R\lambda}$ ($\Sigma=\Sigma_{flux} +\Sigma_{R\lambda}$).
(a) Imaginary part of $\Sigma_{flux}$ at $T=0$ for several dopings 
approaching the QCP $\delta_c\sim 0.23$ for the antinodal Fermi wave vector $k_F^{AN}$. 
(b) The same as (a) for $\Sigma_{R\lambda}$.
Inset (i) shows the quasiparticle weight $Z$ vs. doping for both contributions
and for  $k_F^{AN}$ and $k_F^N$.
While $Z$ is weakly independent of doping and isotropic on the Fermi surface
for $\Sigma_{R\lambda}$, for $\Sigma_{flux}$ $Z$ is strongly anisotropic on 
the FS and strongly doping dependent. Note that $Z_{flux}$ for 
$k_F^{AN}$ tends to zero approaching the QCP.
In addition, at large doping, the relevant contribution is $\Sigma_{R\lambda}$.
Inset (ii) shows, for $\delta=0.24$ and $k=k^{AN}_F$, the $\rm-Im \Sigma_{R\lambda}$ in an extended $\omega$ 
scale, showing that the energy scale in $\Sigma_{R\lambda}$ is large and of 
the order of $t$.
}
\end{figure}
In $\Sigma(k,\omega)$, there is another contribution that is nearly independent of $J$.
This contribution belongs to the usual charge $\delta R$ and nondouble occupancy $\delta \lambda$ channels,
(the first and second components of $h_a$ [Eq.(7)]) and can be written as\cite{greco08,foussats08}
\begin{eqnarray}\label{eq:SigmaIm0Rl}
{\mathrm{Im}} \, \Sigma_{R \lambda}(k,\omega)=
&-& \frac{1}{ N_{s}} \sum_{q} \left\{ \Omega^{2} \; {\mathrm{Im}} [D_{RR}(q,\omega-\epsilon_{{k-q}})] \right. \nonumber \\ 
&+& \;2\;\Omega \; {\mathrm{Im}}[D_{\lambda R}(q,\omega-\epsilon_{{k-q}})] \nonumber \\ 
&+& \left. {\mathrm{Im}} [D_{\lambda \lambda}(q,\omega-\epsilon_{{k-q}})] \right\} \nonumber \\ 
&\times& \left[ n_{F}(-\epsilon_{{k-q}}) + n_{B}(\omega-\epsilon_{{k-q}}) \right],
\end{eqnarray}
where $\Omega=\frac{1}{2}(\epsilon_{{k-q}} + \omega + 2\mu)$.

Figure 2(b) shows $\rm-Im \, \Sigma_{R \lambda}(k,\omega)$ at $T=0$ for $k_F^{AN}$ and
for the same dopings as in Fig. 2(a).
For $k_F^N$, $\rm Im \, \Sigma_{R \lambda}(k,\omega)$ (not shown) is nearly indistinguishable from
results at $k_F^{AN}$, showing that $\Sigma_{R \lambda}$ is rather isotropic on the FS.
In addition, $\Sigma_{R\lambda}$ behaves as $\sim \omega^2$ at low $\omega$, and 
contrary to $\Sigma_{flux}$, there is no energy scale at low energy.
In inset (i) we show $Z_{R\lambda}^{AN}$ and $Z_{R\lambda}^{N}$. These results show 
that the doping dependence of $\Sigma_{R\lambda}$ is weaker than $\Sigma_{flux}$.
Note that $Z_{R\lambda} \rightarrow 0$ when $\delta \rightarrow 0$.
It is important to note that there are no structures in $\rm Im \, \Sigma_{R \lambda}(k,\omega)$ at low energy,
and the main contribution appears at large energies of the order of $t$ [see inset (ii)].
Note also the strong asymmetry shown by $\Sigma_{R\lambda}$ that arises from nondouble-occupancy effects.\cite{foussats08} 

In summary,
(a) $\Sigma_{flux}$ is strongly $J$ and doping dependent, is strongly anisotropic on the FS, and contributes 
at low energy, and
(b) $\Sigma_{R \lambda}$ is  nearly $J$ and doping independent, is strongly isotropic on the FS, and contributes at large energy.  
Thus, $\Sigma(k,\omega)$ can be written as the addition of two well-decoupled channels.
\begin{eqnarray}\label{eq:SigmaIm0}
{\rm Im} \, \Sigma(k,\omega) = {\rm Im} \, \Sigma_{R \lambda}(k,\omega) + {\rm Im} \, \Sigma_{flux}(k,\omega)
\end{eqnarray}
\noindent

Using the Kramers-Kronig relations, $\mathrm{Re} \Sigma(k,\omega)$ can be determined  from $\mathrm{Im}
\Sigma(k,\omega)$ and  the spectral function
$A(k,\omega)$, computed as usual:
\begin{eqnarray}\label{A}
A(k,\omega)= -\frac{1}{\pi}\frac{{\rm Im}\Sigma(k,\omega)}
{[\omega- \epsilon_{k} - {\rm Re}\Sigma(k,\omega)]^2 + {\rm Im}\Sigma(k,\omega)^2} \nonumber \\
\end{eqnarray}

Before concluding this section it is important to note that at mean-field level the $d$-CDW picture
leads below $T_{FP}$, where the translational symmetry is broken, to four hole pockets with low 
spectral weight in the outer side resembling the FAs.\cite{chakravarty03}
However, as discussed in Ref. \onlinecite{norman07}, this picture has conflicting points when compared with some ARPES data. 
We will show in Sec.III that the inclusion of self-energy effects in proximity to 
the FP instability provides a possible scenario for describing several ARPES features. Therefore, although at mean-field level the 
instability to the static $d$-CDW occurs 
below $T_{FP}$, the PG and FA formation do not require the long-range $d$-CDW state, but they do require the enhancement of
fluctuations due to proximity effects. Thus, we are always located in a homogeneous state with the presence 
of $d$-CDW fluctuations.

\section{$\Sigma_{flux}$ and Fermi arcs}

\subsection{Topology of Fermi arcs}

Since $\Sigma_{flux}$ dominates at low energy we study here  
the spectral functions calculated with this contribution.
In Sec. IV we show that the inclusion 
of $\Sigma_{R\lambda}$ does not change the main conclusion obtained in this section. 
\begin{figure}
\centering
\includegraphics[width=8cm]{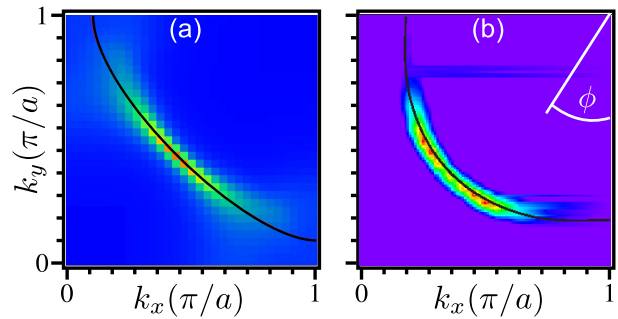}
\label{fig:int}
\caption{(Color online) (a) Intensity of the spectral function at $\omega=0$ vs $k_x,k_y$ for 
$\delta=0.10$ and $T/t=0.025$, above but close to $T_{FP}/t \sim 0.018$. 
(b) The same as (a) but taken from the experimental results of Ref. \onlinecite{norman07}  for comparison.
As in the experiment, (a) shows a well-defined Fermi arc whose end  
does not turn away from the underlying FS (solid line).
}
\end{figure}
Figure 3(a) shows for $\delta=0.10$ and $T/t=0.025$ (above but close to $T_{FP}$) the spectral function 
intensity at zero energy vs $k_x,k_y$.
A well-defined FA is obtained.
Similar to the experiment\cite{norman07} [Fig. 3(b)], the end of the arc does not turn away from the underlying FS, 
and there is no strong suppression of the intensity at the hot spots.

In Fig. 4(a) the intensity of the spectral function on the FS is plotted as a function of the FS angle $\phi$ [defined in Fig. 3(b)]
from the antinode ($\phi=0\circ$) to the node ($\phi=45\circ$).
As in the experiment\cite{norman07} [Fig. 4(b)], the intensity monotonically decreases approaching the
antinode but remains finite.
\begin{figure}
\centering
\includegraphics[width=8cm]{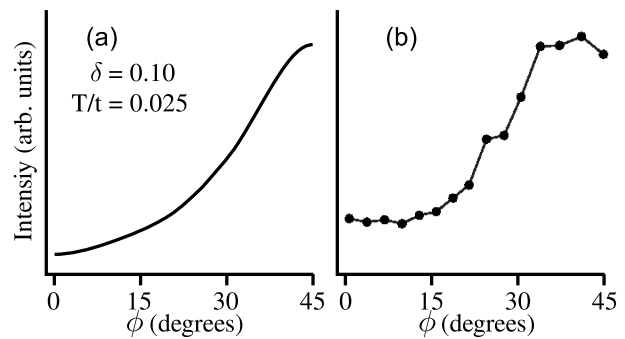}
\caption{(Color online) (a) Intensity of the spectral function at the FS vs the Fermi surface 
angle $\phi$ from the antinode ($\phi=0\circ$) to the node ($\phi=45\circ$)   
for $\delta=0.10$ and $T/t=0.025$.   
(b) The same as (a) but taken from the experimental results of Ref. \onlinecite{norman07} for comparison.
}
\label{fig:angInt}
\end{figure}
In Fig. 5(a) energy distribution curves (EDC) on the underlying FS are plotted.
In agreement with the experiment\cite{norman07} [Fig. 5(b)], near the node, there are well-defined
QP peaks; approaching the antinode,
the spectral functions lose intensity at $\omega=0$, become broad, and develop a PG-like feature.
The presence of a PG-like feature near the antinode means that the 
arc plotted in Fig. 3(a) is not simply related to a decrease 
in the intensity from the node to the antinode but that the FS near the 
antinode is gapped.

Note that the present PG-like feature is not related to a true gap as in other models.
It is developed dynamically in proximity to the $d$-CDW instability in the presence of
strong and short-range $d$-CDW fluctuations.

In summary, the effects described in Figs. 3–5 arise from self-energy effects due to the coupling between QPs 
and the soft flux mode (see inset in Fig. 1) 
in proximity to the FP instability (solid line in Fig. 1).
Since the flux mode occurs mainly with momentum $(\pi,\pi)$, 
the QP near the antinode is distorted, leading to FAs.
Note also that since $d$-CDW fluctuations are of short-range character, in the present picture, the translational
symmetry is not broken.
\begin{figure}
\centering
\includegraphics[width=8cm]{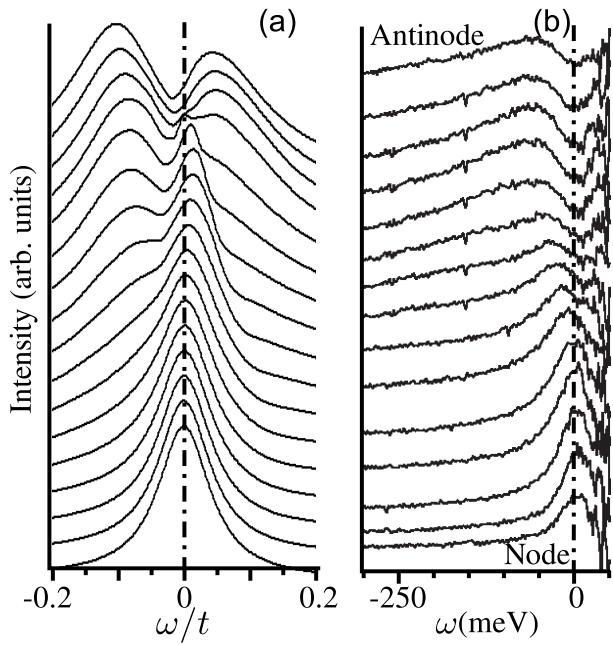}
\label{fig:EDC}
\caption{(a) Energy distribution curves 
on the underlying FS from the antinode (top) to the node (bottom) for $\delta=0.10$ and 
$T/t=0.025$. Near the node, well-defined QP peaks are observed. Moving from the node to the antinode, 
the intensity at $\omega=0$ decreases, the peak becomes broad, and a pseudogap feature is developed.
(b) The same as (a) but taken from the experimental results of Ref. \onlinecite{norman07} for comparison.
}
\end{figure}

\subsection{Temperature dependence of the Fermi arcs}

The temperature dependence of the length of the FAs is puzzling. In spite of different views and interpretations
most reports agree on the fact that the observed FAs depend on
temperature.
While there are reports that claim that the length of FAs collapse to one
isolated point\cite{kanigel06,kanigel07} (nodal metal) at $T=0$, others suggest a less-strong
temperature dependence.\cite{jonson3,storey08}
In models discussed in Ref. \onlinecite{norman07} the temperature dependence of the
length of the arcs arises after assuming a temperature dependence
for $\Delta_k$ or for the lifetime broadening $\Gamma$.
In this subsection it is shown that the temperature dependence of the FAs emerges, 
in the framework of the present approach, without adjustable parameters,
showing that the temperature dependence of the length of the arcs is entangled to their origin.

Figure 6 shows the plot, for $\delta=0.10$, of FA for different temperatures.
Clearly, the length of the arcs decreases when temperature decreases.
We note that the temperature dependence of the arcs seems to be weaker
than in some experiments\cite{kanigel06} but closer to others\cite{jonson3,storey08}
(this point is further discussed in Sec. IV).
Beyond a quantitative comparison, the because no phenomenological parameter is assumed to
be temperature dependent in the present model, the results can be considered satisfactory.
Figure 7 plots the spectral function intensity on the FS for several temperatures normalized to the intensity 
at $k_F^N$.
Consistent with the picture in Fig. 6, 
with decreasing temperature, the intensity is more concentrated around  the node. 

\begin{figure}
 \centering
 \includegraphics[width=8cm]{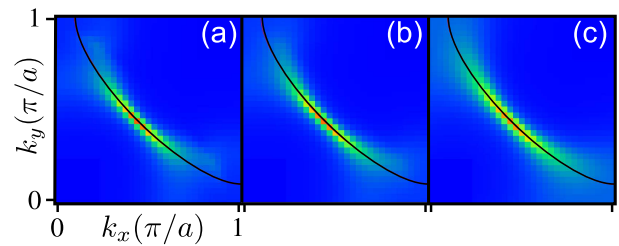}
 \label{fig:int21}
 \caption{(Color online) Fermi arc for $\delta=0.10$ for several temperatures: 
(a) $T/t=0.021$, (b) $T/t=0.025$, and (c) $T/t=0.050$.
Similar to experiments, when the temperature increases, the length of the arc increases. 
}
\end{figure}
\begin{figure}
\centering
\includegraphics[width=7.5cm]{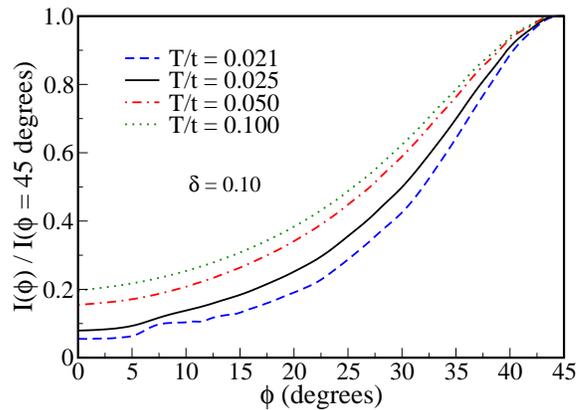}
\caption{(Color online) (a) Intensity of the spectral function (normalized to the 
intensity at the node ) at the FS vs. $\phi$  
for $\delta=0.10$ and for several temperatures. When the temperature decreases toward
$T_{FP}$, the intensity is more and more concentrated around the node.
}
\label{fig:angIntNorm}
\end{figure}
\begin{figure}
\centering
\includegraphics[bb=24 32 706 570,width=7.5cm]{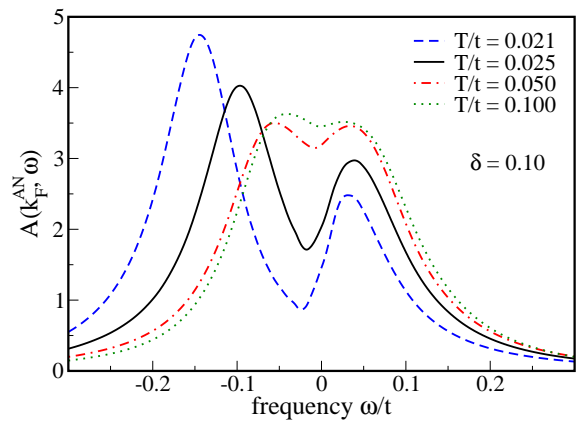}
\caption{(Color online) EDC at $k_F^{AN}$ for $\delta=0.10$ and for the same temperatures as in Fig. 7. As in experiments 
when temperature increases, the PG feature fades out. 
Although the leading edge of the pseudogap partially closes, a filling is also observed, as in the experiments.
}
 \label{fig:AkkT}
\end{figure}

Finally, Fig. 8 shows EDC at $k_F^{AN}$ for the same temperatures as in Fig. 7. 
Although the PG feature partially closes\cite{hashimoto10} with increasing temperature,
a filling is also observed.\cite{norman98,kanigel06,kanigel07}
This feature is in agreement with experiments and in contrast to results from mean-field
calculations where only a closure is expected.

It is worth mentioning that while the arcs discussed here are dynamically generated, they necessarily occur at 
finite temperature. 
The present approach should be distinguished from other ones\cite{kampf,lorenzana,choi} where a phenomenological 
fitted susceptibility without explicit temperature dependence is proposed.

\subsection{Doping dependence of the Fermi arcs}

It is well known that with increasing doping, the PG feature closes\cite{kim98,kaminski05}
and, simultaneously, the length of the arcs increases.\cite{kanigel07}
For describing this behavior, models discussed in Ref. \onlinecite{norman07} need to assume a phenomenological doping dependence for the PG.
In this subsection we will show that arcs whose length increases with increasing doping can be  
naturally discussed in the present context.

In Fig. 9 the FA is shown  for several dopings and for a fixed temperature
$T/t=0.025$. With increasing doping, the length of the arcs increases, in agreement with experiments.
Figure 10  shows EDC at $k_F^{AN}$ for several dopings.
When doping increases, the PG-like feature closes and, simultaneously, the intensity increases at $\omega=0$.

\begin{figure}
\includegraphics[width=8cm]{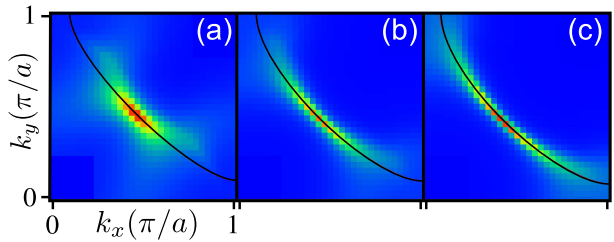}
\caption{(Color online) Fermi arc for $T/t=0.025$ and for several dopings:  
(a) $\delta=0.05$, (b) $\delta=0.10$, and (c) $\delta=0.15$.
As in experiments when doping increases toward overdoped, the length of the arc increases.
}
\label{fig:intdp}
\end{figure}
\begin{figure}
\centering
\includegraphics[bb=24 50 706 570,width=8cm]{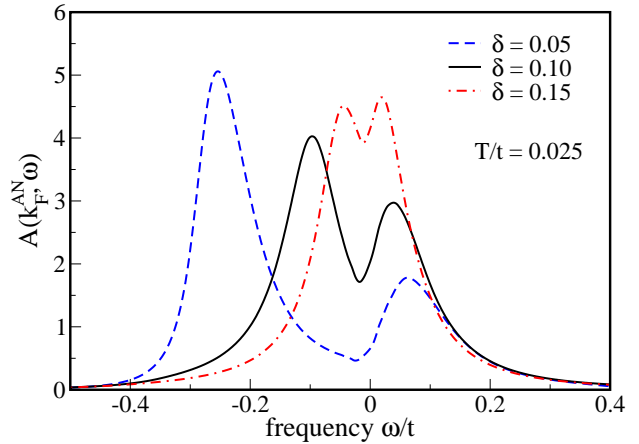}
\caption{(Color online) EDC at $k_F^{AN}$ for $T/t=0.025$ and for the same dopings as in Fig. 9.
Like in experiments when doping increases, the PG feature washes out in a way consistent with the increment of 
the length of the arc reported in Fig. 9.
}
\label{fig:Akdp}
\end{figure}
In summary, in Secs. III B and III C it is shown that with increasing doping and temperature the PG-like
feature and the FA fade out like in the experiments.
The origin for this behavior is easy to understand: By increasing doping and temperature we leave out the instability line $T_{FP}$.
Then, the flux mode is less efficient, self-energy effects become weaker, and the long FS is smoothly recovered.
It is important to note that 
from our approach $T^*$ must be distinguished from a true phase transition.
Here at $T^* > T_{FP}$, where the PG features vanish, there is not a phase transition but a smooth crossover.\cite{tallon01} 
Finally, note that if $t=0.4 eV$, the energy scale for the pseudogap and temperature
is of the order of the experiment.

\subsection{Main characteristics of $\Sigma_{flux}$}

For a complete discussion about the origin of the arcs we have investigated the main characteristics of $\Sigma_{flux}$. 
Figure 11 shows $\rm-Im \Sigma_{flux}$ at $k_F^N$ and $k_F^{AN}$ for $T/t=0.025$ and 
$\delta=0.10$ [Fig. 11(a)], $\delta=0.25$ [Fig. 11(b)], and $\delta=0.40$ [Fig. 11(c)].
At $k_F^N$, $\rm-Im \Sigma_{flux}$ is smaller than for $k_F^{AN}$,
leading to a well-defined and nearly no renormalized QP peak in the nodal direction [Fig. 5(a)]. 
However, the behavior at $k_F^{AN}$ is very different, especially at low doping.
Instead of a minimum at $\omega = 0$, $\rm-Im \Sigma_{flux}$ shows a maximum clearly observed
for $\delta=0.10$ [Fig. 11(a)]. 
This behavior, which is in contrast to the expected results from the usual many-body physics,\cite{katanin03,dellanna06} 
is the main reason for the PG and FA formation. 
With increasing doping, the maximum at $\omega \sim 0$ washes out, and for large doping,
$\rm-Im \Sigma_{flux}$ develops the expected minimum at $\omega = 0$. [See results for $\delta=0.40$ in Fig. 11(c)].

\begin{figure}
\centering
\includegraphics[width=8cm]{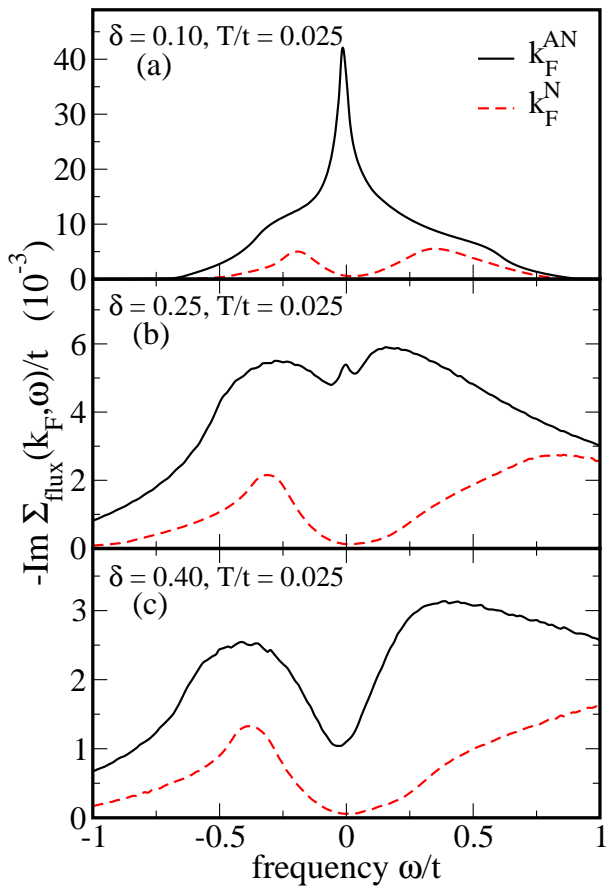}
\label{fig:int100}
\caption{(Color online)
$\rm-Im \Sigma_{flux}$ for $T/t = 0.025$ at $k_F^{AN}$ and $k_F^N$ for (a) $\delta=0.10$, (b) $\delta=0.25$,
and (c) $\delta=0.40$.
For all dopings, $\rm-Im \Sigma_{flux}$  at $k_F^N$ (dashed line) is smaller than the corresponding results
at $k_F^{AN}$ and shows the expected minimum at $\omega=0$.
However, the behavior at $k_F^{AN}$ is very different.
With decreasing doping, $\rm-Im \Sigma_{flux}$ (solid line) increases, and a maximum, instead of a
minimum, is developed at $\omega \sim 0$.
This behavior, which occurs only at finite temperature, is the cause for the dynamical generation of
the arcs and the PG feature. For large doping (see results for $\delta=0.40$),  $\rm-Im \Sigma_{flux}$ is small and depicts the 
expected behavior from the usual many-body theory, i.e., it has a minimum at $\omega=0$.
This behavior is consistent with the fact that no arcs and no PG features are obtained for large doping.
}
\end{figure}

\section{Inclusion of $\Sigma_{R\lambda}$}

\begin{figure}
\centering
\includegraphics[width=8.5cm]{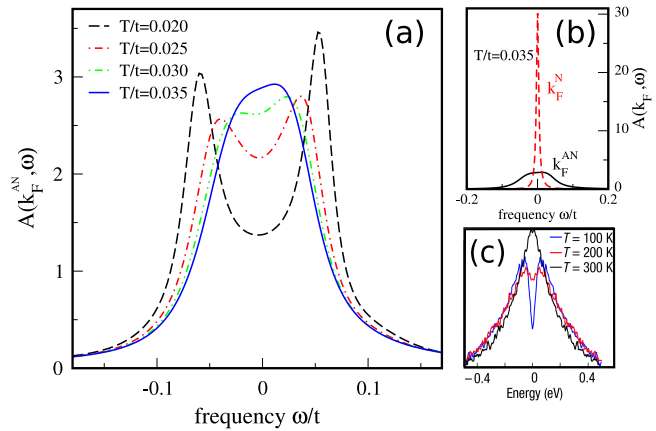}
\label{fig:llena}
\caption{(Color online) (a) EDC at $k_F^{AN}$ for $\delta=0.10$ and for several temperatures 
calculated using  both contributions, $\Sigma_{R\lambda}$ and $\Sigma_{flux}$. 
Note that the inclusion of $\Sigma_{R \lambda}$ does not change the main conclusion
obtained when only $\Sigma_{flux}$ is considered. 
As in the experiments, when temperature increases, the PG feature washes out. 
As in Fig. 8, although the leading edge of the pseudogap closes, a filling is also observed. 
Note that different from the calculation with only $\Sigma_{flux}$ (Fig. 8), at $T/t=0.035$
a full peak is recovered.  
(b) Spectral functions at $k_F^{AN}$ and $k_F^N$ for $T/t=0.035$.
(c) Experimental results taken from Ref. \onlinecite{kanigel06} for comparison.
}
\end{figure}
It was shown (Fig. 2) that the energy scale in $\Sigma_{R\lambda}$ is much larger ($\sim t$) than the energy scale in $\Sigma_{flux}$.
Although this fact implies (as shown in Sec. III) that $\Sigma_{flux}$ is the relevant contribution for triggering
the low-energy PG features, in this section, 
for completeness, we discuss the role of $\Sigma_{R\lambda}$ in the spectral functions.
It was discussed in Sec. IIIB that the PG closes and fills smoothly with increasing temperature (see Fig. 8).
In this section we show that the only role of including $\Sigma_{R \lambda}$ is to improve the vanishing
of the PG.\cite{add}

Figure 12(a) shows EDC for $\delta=0.10$ for several temperatures at $k_F^{AN}$.
This figure shows that with increasing temperature the PG fills and a peak 
at $\omega=0$ emerges at $T/t \sim 0.035$.
Note that in Fig. 8, where only $\Sigma_{flux}$ was considered, even at the
high temperature $T/t =0.1$ the maximum of $A(k,\omega)$ is not yet fully formed at $\omega=0$.
In Fig. 12(c) we have reproduced, for comparison, the experimental results,\cite{kanigel06} showing 
qualitative agreement between theory and experiment.
In Fig. 12(b)  we plot, for $T/t=0.035$, $A(k,\omega)$ for $k_F^{AN}$
(solid line) and $k_F^N$ (dashed line).
Althought the entire FS is ungapped at this temperature, the QP are better defined near the node, as 
in the experiment.\cite{kim98}

\begin{figure}
\centering
\includegraphics[width=8cm]{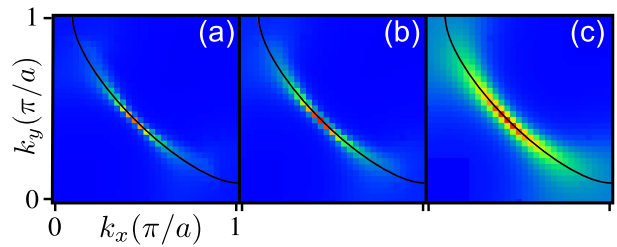}
\label{fig:intRL}
\caption{(Color online) Fermi arc for the same temperatures as in Fig. 6, 
(a) $T/t = 0.021$, (b) $T/t = 0.025$, and (c) $T/t = 0.050$,
but calculated using both $\Sigma_{R\lambda}$ and  $\Sigma_{flux}$. 
As in Fig. 6, when temperature  increases the length of the arc increases. 
}
\end{figure}
Figure 13 shows the spectral function intensity at $\omega=0$ vs $k_x$,$k_y$ for $\delta=0.10$ and
for the same temperatures as in Fig. 6.
At low temperatures a FA is obtained, and its length increases with increasing $T$.
Note that while the FS is expected to be gapped near the antinode for $T/t=0.025$ 
(dot-dashed line in Fig. 12), for $T/t=0.050$ the full FS is ungapped.

In summary, $\Sigma_{R\lambda}$ does not modify the 
main conclusion obtained in Sec. III.
We have shown that its inclusion enhances the pseudogap closing and filling,
and contributes to a faster reconstruction of the entire FS with increasing temperature.

\section{Discussion and conclusion}

The large-$N$ approach in the $t$-$J$ model leads, beyond the mean-field level,  
to two distinct dynamical self-energy contributions, namely, $\Sigma_{R\lambda}$ and $\Sigma_{flux}$. 
In this paper we have analyzed the role of these contributions in ARPES.

The main characteristics of $\Sigma_{flux}$ are the following.
$\Sigma_{flux}$ is strongly anisotropic on the FS, strongly doping dependent, 
and dominated by $J$ (if $J=0$ $\Sigma_{flux}$ is negligible), 
and it contributes at low energy. Thus, $\Sigma_{flux}$ is the relevant 
contribution for describing the Fermi arcs and pseudogap features.

The fact that $\Sigma_{flux}$ is mainly dominated by $J$ may be 
understood as follows. 
At mean-field level the $t$-$J$ model shows (and only 
for finite $J$) the flux or $d$-CDW phase  
below a temperature $T_{FP}$. $T_{FP}$  decreases with increasing doping,
approaching the QCP at $\delta_c$ and $T=0$. 
In the proximity of $T_{FP}$, $d$-CDW fluctuations enter $\Sigma_{flux}$ [Eq.(12)]. 
Since $d$-CDW fluctuations 
favor scattering between electrons with momentum transfer 
$q \sim (\pi,\pi)$, the FS near the antinode is gaped, leading to
Fermi arcs being dynamically generated.
With increasing doping and temperature beyond $\delta_c$ 
and $T_{FP}$, respectively, $d$-CDW fluctuations become weak, and the Fermi arcs and the pseudogap 
wash out, in agreement with experiments.

It is important to note that the
present picture does not require any phenomenological parametrization for 
the pseudogap or the lifetime broadening and their temperature and doping dependence.
It is only necessary to be in the proximity of the $d$-CDW instability
or in a situation with strong short-range fluctuations.
In other words, under the present approach Fermi arcs originate dynamically due to the
interaction between carriers and short-range and short-living 
$d$-CDW fluctuations, implying that long-range order is not broken.

The present picture has similarities with some phenomenological
approaches\cite{choi} where the pseudogap and Fermi arcs are described in a scenario
where fermions interact with bosonic fluctuations of some special order.
Importantly, our description offers a microscopic derivation 
from the $t$-$J$ model, and, as a corollary, the fluctuating spectrum is obtained 
with no assumptions of any fitted phenomenological parameter, such as
coupling, correlation length, or bosonic frequency.
Note that near the flux instability the 
flux mode (inset in Fig. 1) is overdamped and intrinsically temperature dependent and can not be easily 
considered as an Einstein mode as in other approaches.\cite{kampf,lorenzana,choi}

A recent ARPES experiment\cite{hashimoto10} suggests a similar scenario to that 
presented here, i.e., density wave fluctuations without long-range order.
As in that experiment, in our theory, the existence (and persistence with decreasing temperature) of broad and gapped 
spectral features near the antinode means that 
we are not sitting in a phase with long-range order.
It is worth mentioning that under the present approach, below the mean-field temperature $T_{FP}$
the long-range $d$-CDW order occurs; that is, a true gap is formed, and sharp spectral features are expected 
with the corresponding reconstruction of the FS in the form of pockets.\cite{greco09}
From an experimental point of view the existence of long-range order in underdoped cuprates is controversial
and is tied to the following facts.
(a) Some ARPES experiments show well-defined spectral peaks near the antinode in the superconducting state,
while others show broad structures (see Ref. \onlinecite{vishik} and references therein).
(b) While some experiments support the existence of a second order parameter, distinct from but coexisting (and competing)
with superconductivity,\cite{terashima07,hashimoto10,yoshida09,kondo07,kondo09,lee07,tanaka06,ma08}
others claim to observe only one gap feature.\cite{norman98,kanigel06,kanigel07,shi}
(c) While quantum oscillations\cite{QOs} and some ARPES experiments show a reconstruction of the FS 
in the form of pockets,\cite{Nd,meng,jonson1,jonson2,jonson3} other reports show 
only arcs.\cite{kanigel06,kanigel07,norman07} 
Although it is not our aim to solve these puzzles (which requires more theoretical and experimental work), 
we have shown that several aspects related to the Fermi arc phenomenology
can be explained by $d$-CDW proximity effects, showing that this picture is not necessarily 
inconsistent with the notion of arcs.

The characteristics of $\Sigma_{R\lambda}$ are very different from those of $\Sigma_{flux}$.
$\Sigma_{R\lambda}$ is dominated by the usual charge channel and (nearly) independent of $J$.
Thus, this is the relevant contribution for the $J=0$ case.
In addition, it is strongly asymmetric in $\omega$ around
the FS due to nondouble-occupancy effects, rather isotropic on the FS, and rather constant as a function of doping 
(for low to intermediate doping).\cite{foussats08}
Finally, it contributes at large energy of the order of $t$.
Although $\Sigma_{R\lambda}$ is not responsible for the pseudogap and 
Fermi arc formation, it gives an additional contribution to the temperature vanishing of the pseudogap.

$\Sigma_{R\lambda}$ and $\Sigma_{flux}$ may also play a role in 
describing other experiments in cuprates.
(a) Since $\Sigma_{R\lambda}$ shows  high- energy contributions, 
it leads, in the spectral functions, to incoherent 
structures at high binding energy, which offers a possible explanation\cite{foussats08,greco07} 
for  the high-energy features or waterfall effects observed in cuprates.\cite{xie07,meevasana07,graf07,zhang08}
Other theoretical\cite{chinos,zemljic} and experimental\cite{zhang08} reports show a similar conclusion.
(b) The existence of two self-energy contributions 
is also consistent with recent angle-dependent magnetoresistance (ADMR) experiments.\cite{abdel06,abdel07,french09}
These experiments show two different inelastic scattering rates with similar characteristics 
to the self-energy behavior discussed here, i.e., a strongly-doping-dependent 
and anisotropic scattering rate on the FS and another one that is weakly doping dependent and isotropic 
on the FS.
Recently, ADMR experiments were discussed in the context of the present approach.\cite{buzon10}

Here we want to comment on the recent progress on DCA. As discussed in Sec. I DCA shows 
the presence of a pseudogap\cite{huscroft01,macridin06} and Fermi arcs.\cite{parcollet04,werner09,lin10} 
We wish to mention here the similarities 
between our results and those in DCA. For instance, the pole feature at $\omega\sim 0$ and near the antinode that occurs 
in $Im \Sigma$ (Fig. 11), which diminishes with increasing temperature and doping, 
is in remarkable agreement with similar results discussed in Ref. \onlinecite{lin10}. This behavior for the self-energy
leads also to a similar doping and temperature dependence for the PG and FAs. Note that in Ref. \onlinecite{lin10}
the temperature filling of the PG as discussed in the present paperwas also obtained. 
We note again that our results do not require the static long-range order 
of the $d$-CDW. What is needed is the enhancement of the $d$-CDW susceptibility due to fluctuations, as can be seen in the inset of Fig. 1. 
Interestingly, although the static $d$-CDW state was not found in Ref. \onlinecite{macridin04}, an enhancement of the $d$-CDW susceptibility 
was obtained. 
Finally, it is worth mentioning that although the origin of the PG and FAs is presumably of antiferromagnetic nature,\cite{huscroft01,macridin06}
a recent report\cite{lin10} is not conclusive about this affirmation. 
It is the aim of the present paper to show that $d$-CDW fluctuations may contribute to the PG and FAs formation.   

Although of one could certainly suppose that the large-$N$ is a particular 
approximation and some results may depend on its details, we think that our theory contains 
features that can be expected, qualitatively, in cuprates and in the $t$-$J$ model. 
Since the low-energy pseudogap feature increases with decreasing doping, it is reasonable to think that 
the pseudogap is associated with the same energy scale as the antiferromagnetism, i.e., $J$. 
This fact is  contained in $\Sigma_{flux}$. On the other hand,
there is a larger energy scale, the hopping $t$, which, together with 
nondouble-occupancy effects, enters through $\Sigma_{R\lambda}$.

Finally, it is worth mentioning that besides $d$-CDW, other
instabilities like stripes,\cite{stripes} antiferromagnetism,\cite{AF} and Pomeranchuk\cite{pomeranchuk}
have been proposed to exist at low doping in cuprates.
Thus, it is important to perform similar calculations for those cases and compare different predictions.

\noindent{\bf Acknowledgments}

The authors thank H. Parent for suggestions  on the presentation of the paper. A.G. thanks R.-H. He, 
W. Metzner, A. Muramatsu, Y. Yamase, and R. Zeyher for valuable discussions and the Max Planck Institute (Stuttgart) and the University 
of Stuttgart for hospitality.


\begin{thebibliography}{10}
\bibitem{add}
After a comparison with the experiments, we have found that a reduction of the contribution
of $\Sigma_{R\lambda}$ by a factor of four is appropiated.
A possible reason may stem from the fact that the large-$N$ approach overstimates 
charge fluctuations over other fluctuations such as spin excitations (see Ref.[\onlinecite{foussats04}] for discussions).
\end{thebibliography}
\end{document}